\documentclass{svjour2}                    
\usepackage{amsmath,amssymb}
\usepackage{graphicx}
\usepackage{psfrag}
\journalname{General Relativity and Gravitation}
\begin{document}

\title{On the generalized Jacobi equation}
\author{Volker Perlick}
\institute{V. Perlick \at
              Physics Department \\
              Lancaster University \\
              Lancaster LA1 4YB, United Kingdom\\
              \email{v.perlick@lancaster.ac.uk}          
}

\date{Received: date / Accepted: date}

\maketitle

\begin{abstract}
The standard text-book Jacobi equation (equation of geodesic deviation)
arises by linearizing the geodesic equation around some chosen geodesic,
where the linearization is done with respect to the coordinates and the
velocities. The generalized Jacobi equation, introduced by Hodgkinson
in 1972 and further developed by Mashhoon and others, arises if the
linearization is done only with respect to the coordinates, but not with
respect to the velocities. The resulting equation has been studied by
several authors in some detail for timelike geodesics in a Lorentzian 
manifold. Here we begin by briefly considering the generalized Jacobi 
equation on affine manifolds, without a metric; then we specify to 
lightlike geodesics in a Lorentzian manifold. We illustrate the latter 
case by considering particular lightlike geodesics (a) in Schwarzschild 
spacetime and (b) in a plane-wave spacetime.
\end{abstract}

\section{Introduction}
\label{intro}
The Jacobi equation, also known as the equation of geodesic deviation, 
describes geodesics in the neighborhood of a reference geodesic. More 
precisely, the Jacobi equation is the linearization of the geodesic 
equation around the reference geodesic, where the linearization is 
done with respect to the coordinates and the velocities. Thus, the
Jacobi equation describes geodesics which are close to the reference 
geodesic and whose velocities are close to the velocity of the reference 
geodesic. The Jacobi equation has important applications to General
Relativity: For timelike geodesics, the Jacobi equation describes
the relative acceleration of freely falling particles around a 
freely falling reference particle. This relative acceleration, which
is determined by the curvature tensor along the worldline of the 
reference particle, can be interpreted as the tidal force produced by 
the gravitational field. For lightlike geodesics, the Jacobi equation 
determines the shape of light bundles around a reference light ray. 
Again, the expansion and distortion of the bundle is determined 
by the curvature tensor along the reference light ray. Detailed 
discussions of the Jacobi equation (or geodesic deviation equation)
can be found in almost any text-book on General Relativity, see 
e.g. Synge \cite{Synge1960}, Misner, Thorne and 
Wheeler \cite{MisnerThorneWheeler1973}
or Hawking and Ellis \cite{HawkingEllis1973}.

If a geodesic is close to a reference geodesic but its velocity is not, 
the Jacobi equation does not give a valid approximation for it. Such a
geodesic is properly described by an equation which is linearized only 
with respect to the coordinates but not with respect to the velocities. 
Such a generalized Jacobi equation was brought forward, for timelike 
geodesics in a general-relativistic spacetime, by 
Hodgkinson \cite{Hodgkinson1972}. This equation describes tidal forces 
on particles whose relative velocity is not small. Applications to 
astrophysics were given by Mashhoon \cite{Mashhoon1975,Mashhoon1977}. 
The generalized Jacobi equation was independently rediscovered by
Ciufolini \cite{Ciufolini1986}. More recently, its implications 
were studied in a series of papers by Chicone and Mashhoon
\cite{ChiconeMashhoon2002,ChiconeMashhoon2005,ChiconeMashhoon2006}.

In geometric terms, the ordinary Jacobi equation approximates
the geo\-desic flow on a tubular neighborhood of the reference
geodesic in phase space (i.e., in the cotangent bundle over
the base manifold). By contrast, the generalized Jacobi 
equation approximates the geodesic flow on a neighborhood in
phase space that is unrestricted in the fiber dimension.

All the references quoted above consider the generalized Jacobi
equation for timelike geodesics, which is physically of particular
relevance because of its relation to tidal forces. In this article 
we want to discuss the generalized Jacobi equation for lightlike 
geodesics. Then it describes light rays around a reference light
ray. The domain where the generalized Jacobi equation is applicable
but the Jacobi equation is not encompasses all neighboring light rays
that are close to the reference light ray but whose velocities
are not. 

Although we are mainly interested in lightlike geodesics, it
is worthwile to note that the geodesic equation and, hence, the 
Jacobi equation and the generalized Jacobi equation can be 
formulated on an affine manifold, without a metric. All that 
is needed is a connection. We begin by deriving the affine 
generalized Jacobi equation in Section \ref{sec:affine}. In
Section \ref{sec:lightlike} we specify to lightlike geodesics
of a Lorentzian metric. In this case, the generalized Jacobi 
equation can be used to describe light bundles with arbitrarily 
large opening angle around a reference light ray. Of course, if 
the opening angle is large, the light rays will leave the neighborhood
of the reference light ray soon, unless they are being refocused 
towards the reference light ray. So the generalized Jacobi equation 
can be used, in general, as a valid approximation only close to 
the vertex of the bundle. The general results of this section are 
illustrated by considering a bundle around a circular light ray 
at $r=3m$ in Schwarzschild spacetime. In Section \ref{sec:Fermi} 
we consider lightlike Fermi normal coordinates which allow to 
write the generalized Jacobi equation in terms of the curvature 
tensor along the reference light ray. As an example, we consider 
a bundle of light rays in a plane-wave spacetime.

Throughout we work with the generalized Jacobi equation in
the sense of Hodgkinson, Mashhoon etc. This equation is 
linearized with respect to the coordinates but retains 
full dependence of the velocities. As an alternative, one 
may approximate with respect to the coordinates and the
velocities up to common order $N$. This approach, which 
dates back to Ba{\.z}a{\'n}ski \cite{Bazanski1977}, can be
used to set up an iterative scheme. One begins by solving
the ordinary Jacobi equation, then calculates the corrections
due to second-order terms and so on. This method was applied
to particle motion in Schwarzschild and Kerr 
spacetimes by Kerner et al. 
\cite{KernerHoltenColistete2001,ColisteteLeygnacKerner2002}.

This paper is dedicated to Bahram Mashhoon on occasion of his
60th birthday. I have very much profited from discussions with
him and from reading his papers. In particular, I wish to thank
him for introducing me to the subject of the generalized
Jacobi equation. 

\section{The generalized Jacobi equation in an affine manifold}
\label{sec:affine}

An affine manifold is a manifold with an affine connection $\nabla$.
In local coordinates, the connection is determined by its connection
coefficients $\Gamma ^{\mu} _{\nu \sigma}$, 
\begin{equation}\label{eq:Gamma}
\nabla _{\partial _{\nu}} \partial _{\sigma} \, = \, 
\Gamma ^{\mu} _{\nu \sigma} \, \partial _{\mu} \, .
\end{equation}
Here and in the following, the dimension of the manifold is $n$,
and we use the summation convention for greek indices running 
from 1 to $n$.

Whenever we have an affine manifold, we can consider the geodesic
equation
\begin{equation}\label{eq:geo} 
\frac{d^2 x^{\mu}}{ds^2} \, + \, \Gamma ^{\mu} _{\nu \sigma} (x) \,
\frac{d x^{\nu}}{ds}  \, \frac{d x^{\sigma}}{ds}  \, =  \, 0 \, .
\end{equation}
The solution curves $x^{\mu} (s)$ of (\ref{eq:geo}) are the
geodesics (autoparallels) of the affine connection. As a geodesic
remains a geodesic under affine reparametrization, $s \mapsto 
as + b$, it is usual to refer to $s$ as to an ``affine parameter''. 
Note that only the symmetric part of $\Gamma ^{\mu} _{\nu \sigma}$
enters into the geodesic equation. The antisymmetric part 
\begin{equation}\label{eq:torsion}
T^{\mu} _{\nu \sigma} \, := \, 
\Gamma ^{\mu} _{\nu \sigma} 
\, - \,  
\Gamma ^{\mu} _{\sigma \nu} \, ,
\end{equation}
which is called the ``torsion'', drops out from the geodesic 
equation. As in the following we are interested only in the 
geodesic equation, and in equations derived 
thereof, we may replace any connection by its symmetrized 
version. Therefore we will assume in the following that the 
connection is symmetric, $\Gamma ^{\mu} _{\nu \sigma} =
\Gamma ^{\mu} _{\sigma \nu}$. (For the Jacobi equation and
some of its generalizations in terms of a non-symmetric 
connection see, e.g., Swaminarayan and Safko
\cite{SwaminarayanSafko1983}.)

Now fix a geodesic $X^{\mu} (s)$,
\begin{equation}\label{eq:geoX} 
\frac{d^2 X^{\mu}}{ds^2} \, + \, \Gamma ^{\mu} _{\nu \sigma} (X) \,
\frac{d X^{\nu}}{ds}  \, \frac{d X^{\sigma}}{ds}  \, =  \, 0 \, .
\end{equation}
We will call it the ``reference geodesic'',
and we will assume that it is known, i.e.,
that we have the $X^{\mu}$ explicitly as functions 
of the curve parameter $s$. 
Then the geodesic equation for a neighboring curve 
$x^{\mu} (s) \, = \, X^{\mu} (s) \, + \, \xi ^{\mu} (s)$ 
results from inserting  $x^{\mu} (s) \, = \, X^{\mu} (s) \, + \, 
\xi ^{\mu} (s)$ into (\ref{eq:geo}) and subtracting
(\ref{eq:geoX}),
\begin{gather}\label{eq:geoxi} 
\frac{d^2  \xi ^{\mu}}{ds^2} \, + \, \Gamma ^{\mu} _{\nu \sigma} (X+ \xi)
\, \Big( \, \frac{dX ^{\nu}}{ds} + \frac{d \xi ^{\nu}}{ds} \, \Big) 
\, \Big( \, \frac{dX ^{\sigma}}{ds} + \frac{d \xi ^{\sigma}}{ds} \, \Big) 
\\[0.3cm]
\nonumber
 - \; 
\Gamma  ^{\mu} _{\nu \sigma} (X) \, 
\frac{d X ^{\nu}}{ds}  \, \frac{d X ^{\sigma}}{ds} 
\, = \, 0 \, .
\end{gather}
With $X^{\mu} (s)$ known, (\ref{eq:geoxi}) is a system of
second order ordinary differential equations for the $n$
functions $\xi ^{\mu} (s)$. It is the exact geodesic equation,
expressed in terms of the coordinate difference $\xi ^{\mu} (s)$
with respect to the reference geodesic $X ^{\mu} (s)$. 

If we linearize (\ref{eq:geoxi}) with respect to $\xi ^{\mu}$, 
but not with respect to $d \xi ^{\mu} /ds$, we get
\begin{gather}\label{eq:genjac}
\frac{d ^2 \xi ^{\mu}}{ds^2} \, + \, 
\Gamma ^{\mu} _{\nu \sigma} (X) \, 
\Big( \, 
2 \, \frac{d \xi ^{\nu}}{ds} \, \frac{d X ^{\sigma}}{ds} \, + \, 
\frac{d \xi ^{\nu}}{ds} \, \frac{d \xi ^{\sigma}}{ds} 
\, \Big)
\\[0.3cm]
\nonumber
 + \;
\partial _{\tau} \Gamma ^{\mu} _{\nu \sigma} (X) \, \xi ^{\tau} \, 
\Big( \, \frac{d X ^{\nu}}{ds} + \frac{d \xi ^{\nu}}{ds} \, \Big) \, 
\Big( \, \frac{d X ^{\sigma}}{ds} + \frac{d \xi ^{\sigma}}{ds} \, \Big) \,
 = \, 0 \, .
\end{gather}
This is the generalized Jacobi equation, on an affine manifold, in
arbitrary coordinates.

By contrast, if we linearize (\ref{eq:geoxi}) both with respect
to $\xi ^{\mu}$ and with respect to $d \xi ^{\mu}/ds$, we get 
\begin{equation}\label{eq:jac}
\frac{d^2 \xi ^{\mu}}{ds^2} \, + \, 
\Gamma ^{\mu} _{\nu \sigma} (X) \, 
2 \, \frac{d \xi ^{\nu}}{ds} \, \frac{d X ^{\sigma}}{ds} 
\, + \, 
\partial _{\tau} \Gamma ^{\mu} _{\nu \sigma} (X) \, \xi ^{\tau} \, 
\frac{d X ^{\nu}}{ds}  \, 
\frac{d X ^{\sigma}}{ds} \,
 = \, 0 \, .
\end{equation}
This is the ordinary Jacobi equation in arbitrary coordinates. To
recover the standard text-book form, we have to introduce the 
covariant derivative along $X(s)$, which is defined by
\begin{equation}\label{eq:cov}
\frac{D \eta ^{\mu}}{ds}  \, = \, \frac{d \eta ^{\mu}}{ds} \,
+ \Gamma ^{\mu} _{\rho \tau} (X) \, \eta ^{\rho} \,  
\frac{d X^{\tau}}{ds} 
\end{equation}
for any $\eta ^{\nu}$, and the curvature tensor
\begin{equation}\label{eq:curv}
R^{\mu} _{\tau \nu \sigma} \, = \,
\partial _{\nu} \Gamma^{\mu} _{\tau \sigma} \, - \,
\partial _{\tau} \Gamma^{\mu} _{\nu \sigma} \, + \,
\Gamma^{\mu} _{\nu \lambda} \, \Gamma^{\lambda} _{\tau \sigma} \, - \,
\Gamma^{\mu} _{\tau \lambda} \, \Gamma^{\lambda} _{\nu \sigma} \, .
\end{equation}
Then a straight-forward calculation reduces (\ref{eq:jac}) to
\begin{equation}\label{eq:jaccov}
\frac{D^2 \xi ^{\mu}}{ds^2} \, + \, 
R^{\mu} _{\tau \nu \sigma} (X) \, \xi ^{\nu} \, 
\frac{dX^{\tau}}{ds} \, \frac{dX^{\sigma}}{ds} \, = \, 0 \, ,
\end{equation} 
which is, indeed, the standard text-book form of the ordinary
Jacobi equation.

The generalized Jacobi equation (\ref{eq:genjac}) is a second order 
non-linear ordinary differential equation for $\xi ^{\mu} (s)$. It is 
non-autonomous because the coefficients $\Gamma ^{\mu} _{\nu \sigma}(X)$
and $\partial _{\tau} \Gamma ^{\mu} _{\nu \sigma}(X)$ are functions
of the curve parameter $s$. It gives a valid approximation for all
those geodesics for which the $\xi ^{\mu} (s)$ are small, whereas
the $d \xi ^{\mu} (s)/ds$ need not be small. Of course, if the 
$d \xi ^{\mu} (s)/ds$ are not small, the $\xi ^{\mu} (s)$ will 
in general remain small only for a small interval of the parameter $s$.
Such geodesics will leave the neighborhood of the reference 
geodesic soon and the generalized Jacobi equation will, in 
general, give a valid approximation for them only on a 
small interval of the parameter $s$.

The generalized Jacobi equation may be modified in two ways.
\begin{itemize}
\item
One may prefer to choose a different, non-affine, parametrization
for the (approximate) geodesics.
\item
One may wish to choose special coordinates such that the
generalized Jacobi equation takes a simpler form and that 
it can be easier compared for two different geodesics.
\end{itemize}
Both is possible in an affine manifold, without refering to
a metric.

As to the first item, it is often convenient to use on a 
geodesic a (non-affine) parameter $u$ which coincides with
one of the coordinates, say $u=x^n$. This is possible for 
all curves which are transverse to the hypersurfaces 
$x^n = \mathrm{constant}$. Then the dependence of the 
$x^n$ coordinate on the parameter is known, $x^n(u)=u$, 
and the geodesic equation reduces to a system of 
differential equations for the remaining coordinate functions
$\big( x^1(u) , \dots , x^{n-1}(u) \big)$. The condition that
two curves $X(u)$ and $X(u) + \xi (u)$, both parametrized
by $u=x^n$, are geometrically close is equivalent 
to the condition that the $\xi ^i (u)$ are small for
$i = 1 , \dots , (n-1)$. By contrast, if an affine 
parameter $s$ is used it may be that not all of
the $ \xi ^{\mu} (s)$ are small even if the two
geodesics are geometrically close. This happens if
the affine parameter  on one of the two geodesics
lags behind or runs ahead of the other. 

Therefore we will now rewrite the generalized Jacobi 
equation with respect to the new curve parameter $u= x^n$.
We first observe that the $\mu =n$ component
of the geodesic equation (\ref{eq:geo}) can be 
written in the form
\begin{equation}\label{eq:geon}
\frac{d^2 u}{ds^2} \, + \, \Gamma ^{n} _{\nu \sigma} (x) \,
\frac{d x^{\nu}}{du}  \, \frac{d x^{\sigma}}{du}  \,
\Big( \, \frac{du}{ds} \, \Big)^2 \, =  \, 0 \, ,
\end{equation}
and the remaining $(n-1)$ components as 
\begin{equation}\label{eq:geoi}
\frac{d}{ds} \Big( \frac{du}{ds} \frac{d x^{i}}{du} \, \Big) \, + \, 
 \Gamma ^{i} _{\nu \sigma} (x) \,
\frac{d x^{\nu}}{du}  \, \frac{d x^{\sigma}}{du}  \, 
\Big( \, \frac{du}{ds} \, \Big)^2 \, =  \, 0 \, .
\end{equation}
Here and in the following, latin indices $i,j, \dots$ take values 
1 to $(n-1)$. Calculating the first term of (\ref{eq:geoi}) with 
the product rule and inserting (\ref{eq:geon}) results in  
\begin{equation}\label{eq:geored}
\frac{d^2 x^i}{du^2}  \, - \, 
\Gamma ^{n} _{\nu \sigma} (x) \,
\frac{d x^{\nu}}{du}  \, \frac{d x^{\sigma}}{du}  \,
\frac{dxi}{du} \, + \, 
 \Gamma ^{i} _{\nu \sigma} (x) \,
\frac{d x^{\nu}}{du}  \, \frac{d x^{\sigma}}{du}  
\,  =  \, 0 \, .
\end{equation}
Together with the equation $x^n (u) = u$, (\ref{eq:geored}) determines 
the geodesics (or, more precisely, those geodesics that are transverse
to the surfaces $x^n = \mathrm{constant}$) with the non-affine
parametrization by $u = x^n$. (\ref{eq:geored}) is a system of second 
order ordinary differential equations for the $(n-1)$ functions 
$x^i (u)$. Clearly, these differential equations are non-autonomous 
because the coefficients depend on $u=x^n$. Also, as a result of the 
reparametrization we have now got a system of equations that is
cubic, rather than quadratic, in the velocities. 

In analogy to our earlier procedure we search for solutions of 
(\ref{eq:geored}) in the form $x^{\mu} (u) = X ^{\mu} (u) 
+ \xi ^{\mu} (u)$, with $X^{\mu}$ a known geodesic, now reparametrized
by $u = x^n$. Of course, as $x^n (u) = X^n(u)=u$, we have $\xi ^n (u) =0$.
Inserting $x^{i} (u) = X ^{i} (u) + \xi ^{i} (u)$ into (\ref{eq:geored}), 
and linearizing with respect to the $\xi ^{i}$, but not with respect to
the $d \xi ^i /du$, gives us the reparametrized version of the 
generalized Jacobi equation,
\begin{gather}\label{eq:genjacred}
\frac{d^2 \xi ^i}{du^2}  
\, + \,
\Gamma ^{i} _{\nu \sigma} (X) \,  \Big( \, 
2 \, \frac{d X^{\nu}}{du}  \, \frac{d \xi ^{\sigma}}{du}  \, + \, 
\frac{d \xi ^{\nu}}{du}  \, \frac{d \xi ^{\sigma}}{du}  \, \Big)
\\[0.3cm]
\nonumber
 - \, 
\Gamma ^{n} _{\nu \sigma} (X) \,  \Big( \, 
2 \, \frac{d X^{\nu}}{du}  \, \frac{d \xi ^{\sigma}}{du}  \, + \, 
\frac{d \xi ^{\nu}}{du}  \, \frac{d \xi ^{\sigma}}{du}  \, \Big)
\, \frac{dX^i}{du} 
\\[0.3cm]
\nonumber
 - \, 
\Gamma ^{n} _{\nu \sigma} (X) \,  
\Big( \frac{d X^{\nu}}{du}  \, + \, \frac{d \xi ^{\nu}}{du}   \, \Big)
\Big( \frac{d X^{\sigma}}{du}  \, + \, \frac{d \xi ^{\sigma}}{du}   \, \Big)
\, \frac{d \xi ^i}{du}
\\[0.3cm]
\nonumber
 + \, 
\partial _{\tau} \Gamma ^i _{\nu \sigma} (X) \,  \xi ^{\tau} \, \Big( \, 
\Big( \frac{d X^{\nu}}{du}  \, + \, \frac{d \xi ^{\nu}}{du}   \, \Big)
\Big( \frac{d X^{\sigma}}{du}  \, + \, \frac{d \xi ^{\sigma}}{du}   \, \Big)
\\[0.3cm]
\nonumber
 - \, 
\partial _{\tau} \Gamma ^n _{\nu \sigma} (X) \,  \xi ^{\tau} \, \Big( \, 
\Big( \frac{d X^{\nu}}{du}  \, + \, \frac{d \xi ^{\nu}}{du}   \, \Big)
\Big( \frac{d X^{\sigma}}{du}  \, + \, \frac{d \xi ^{\sigma}}{du}   \, \Big)
\Big( \frac{d X^{i}}{du}  \, + \, \frac{d \xi ^{i}}{du}   \, \Big)
\, = \, 0 \, . 
\end{gather}

We now turn to the question of whether the generalized Jacobi equation can
be simplified by choosing special coordinates. To that end we will use
Fermi coordinates. Text-books on general relativity treat Fermi 
coordinates near a timelike reference geodesic (or, more generally,
near a timelike reference curve) in a Lorentzian manifold, see e.g. 
Synge \cite{Synge1960}. However, it is fairly obvious that, by an
analogous procedure, one can introduce Fermi coordinates near an 
arbitrary reference geodesic in an affine manifold, without a 
metric. The construction is as follows.

Let an affinely parametrized geodesic $X(s)$ be given. Choose an $n$-bein 
(i.e., $n$ linearly independent vectors) at one point of the geodesic, 
such that the $n$th vector coincides with the tangent vector of the 
geodesic, and transport this $n$-bein parallelly along $X(s)$.
Denote the $n$ resulting vector fields along the geodesic by $E_1 (s), 
\dots , E_n (s)$. As $X(s)$ is an affinely parametrized geodesic, $E_n (s)$
coincides with the tangent vector $d X(s) /ds$ for all $s$.
Let exp denote the exponential map of the
given connection which, as indicated above, is assumed to be 
torsion-free. Then every point in a sufficiently small tubular 
neighborhood of the given geodesic can be written uniquely 
as $\mathrm{exp} _{X(u)} ( x^i E_i (u) )$. The $n$ numbers $(x^1, \dots , x^{n-1} , u)$
are the Fermi coordinates of this point. By definition of the exponential
map, this means that the Fermi coordinates are determined by the 
following property: The point with Fermi coordinates 
$(x^1, \dots , x^{n-1} , u)$ can be reached by following
the geodesic with initial point $X(u)$ and initial tangent vector 
$x^i E_i (u)$ up to the affine parameter 1. By construction, the Fermi
coordinate $u$ coincides along the geodesic $X(s)$ with the
affine parameter $s$. Along neighboring geodesics, however,
the Fermi coordinate $u$ will give a non-affine parametrization
in general.   

Note that Fermi coordinates are well-defined only on a (possibly small)
tubular neighborhood of a chosen geodesic $X(s)$. Farther away from $X(s)$
they need not exist, because there may be points that cannot be reached
by a geodesic starting on $X(s)$, and they need not be unique, because
geodesics issuing from a point on $X(s)$ may intersect.  

The crucial property of Fermi coordinates is that, in such coordinates,
the connection coefficients $\Gamma ^{\mu} _{\nu \sigma}$ vanish on $X(s)$. 
To prove this, we first observe that $\nabla _{\partial _n} \partial _{\mu} 
= 0$ along $X(s)$ because the $E_{\mu}$ are parallelly transported,
hence $\Gamma ^{\mu} _{n \sigma} (X)= 0$. As the connection is 
symmetric, this implies that $\Gamma ^{\mu} _{\sigma n} (X) 
= 0$. What remains to be proven is that $\Gamma ^{\mu} _{ij} (X)
= 0$. We use the fact that
\begin{equation}\label{eq:ij}
\nabla _{( \partial _i + \partial _j )} 
\big( \partial _i + \partial _j \big) \, = \, 
\nabla _{\partial _i } 
\partial _i  \, + \, 
\nabla _{\partial _j } 
\partial _j  \, + \, 
\nabla _{\partial _i } 
\partial _j  \, + \, 
\nabla _{\partial _j } 
\partial _i  \, . 
\end{equation}
As for any $(n-1)$-tuple $(x^1, \dots , x^{n-1})$ the 
integral curves of $x^i \partial _i$ that start on $X(s)$ are 
geodesics, the left-hand side and the first two terms on the
right-hand side of (\ref{eq:ij}) vanish on $X(s)$. Thus, 
(\ref{eq:ij}) tells us that $\Gamma ^{\mu} _{ij} (X) +  
\Gamma ^{\mu} _{ji} (X) =0$. As our connection is symmetric, 
this implies that, indeed, $\Gamma ^{\mu} _{ij} (X) = 0$.   
 
If written in Fermi coordinates, the generalized Jacobi equation 
(\ref{eq:genjacred}) simplifies considerably because of 
$\Gamma ^{\mu} _{\nu \sigma} (X)=0$. Moreover, the use of 
Fermi coordinates reduces the generalized Jacobi equation to
a standard form which facilitates comparison of this equation
for two different geodesics. In Section \ref{sec:Fermi} below
we will make this explicit for lightlike geodesics in a Lorentzian 
manifold, where the Fermi coordinates can be further specified.

\section{The generalized Jacobi equation for lightlike geodesics in
arbitrary coordinates}
\label{sec:lightlike}
We now specify the results of the preceding section to the case that
$\nabla$ is the Levi-Civita connection of a pseudo-Riemannian metric
$g = g_{\mu \nu} dx^{\mu} dx^{\nu}$,
\begin{equation}\label{eq:LC}
\Gamma ^{\mu} _{\nu \sigma} \, = \, \frac{1}{2} \, g^{\mu \rho} \,
\big( \, \partial _{\nu} g_{\sigma \rho} \, + \, 
\partial _{\sigma} g_{\nu \rho} \, - \, 
\partial _{\rho} g_{\nu \sigma} \, \big) \, ,
\end{equation}
where, as usual,
\begin{equation}\label{eq:gcontra}
g^{\mu \nu} \, g_{\nu \sigma} \, = \, \delta ^{\mu} _{\sigma} \, .
\end{equation}
We assume that the metric has Lorentzian signature $(+, , \dots , + , -)$,
and we want to discuss the generalized Jacobi equation for the case
of lightlike geodesics,
\begin{equation}\label{eq:lightlike}
g_{\mu \nu} \, \frac{dx^{\mu}}{ds} \, \frac{dx^{\nu}}{ds} \, = \, 0 \, .
\end{equation}  
If the dimension $n$ of the manifold is equal to 4, lightlike geodesics
can be interpreted as light rays in a general-relarivistic spacetime.
The mathematical results, however, hold for any $n$. 

Let us briefly recall the well-known fact that, if the connection 
coefficients $\Gamma ^{\mu} _{\nu \sigma}$ are given in terms of a 
metric via (\ref{eq:LC}), the geodesic equation (\ref{eq:geo}) can 
be derived from a Hamiltonian. As a matter of fact, Hamilton's equations 
for the Hamiltonian
\begin{equation}\label{eq:Ham}
H(x,p) \, = \, \frac{1}{2} \, g^{\mu \nu} (x) \, p_{\mu} \, p_{\nu} \, .
\end{equation}
take the form
\begin{equation}\label{eq:Hamx}
\frac{dx^{\mu}}{ds} \, = \, p_{\mu} \, ,
\end{equation}
\begin{equation}\label{eq:Hamp}
\frac{dp_{\mu}}{ds} \, = \, 
\frac{1}{2} \, ( \partial _{\mu} g^{\rho \sigma})(x) \, 
p_{\rho} \, p_{\sigma} \, .
\end{equation}
The derivative with respect to $s$ of (\ref{eq:Hamp}), together
with (\ref{eq:Hamx}) and (\ref{eq:LC}), indeed reproduces the
geodesic equation (\ref{eq:geo}). By (\ref{eq:Hamx}), the 
side-condition (\ref{eq:lightlike}) takes the form
\begin{equation}\label{eq:cone}
H(x,p) \, = \, 0 \, .
\end{equation}  
A reparametrization of the solution curves can be achieved for
lightlike geodesics in a particularly convenient way: We just
have to multiply the Hamiltonian with an appropriate function
of $x$ and $p$. If we want to change from the affine parameter
$s$ to the parameter $u = x^n$, we can proceed in the following
way. We first recall that such a reparametrization is possible
for all geodesics that are transverse to the hypersurfaces $x^n =
\mathrm{constant}$. (This condition is true for \emph{all} 
lightlike geodesics if and only if the hypersurfaces $x^n =
\mathrm{constant}$ are spacelike. However, we may also 
consider the case that these hypersurfaces are lightlike or 
timelike; then the transversality condition only holds for 
\emph{some} lightlike geodesics.) By (\ref{eq:Hamx}) and 
(\ref{eq:lightlike}), the transversality condition is true if 
$g^{n \sigma} (x) p_{\sigma} \neq 0$. This condition
determines an open subset of the cotangent bundle (phase space).
On this subset we can switch from the Hamiltonian (\ref{eq:Ham})
to the modified Hamiltonian 
\begin{equation}\label{eq:Hamred}
\tilde{H} (x,p) \, = \, 
\frac{g^{\mu \nu} (x) \, p_{\mu} \, p_{\nu}}{2 \, g^{n \sigma}(x) \, p_{\sigma}}
\, .
\end{equation} 
Hamilton's equation with the Hamiltonian $\tilde{H}$ read
\begin{equation}\label{eq:tHamx}
\frac{d x^{\mu}}{du} \, = \, - \, 
\frac{\tilde{H}(x,p) \, g^{n \mu} (x)}{g^{n \sigma}(x) \, p_{\sigma}}
\, + \, 
\frac{g^{\mu \nu} (x) \, p_{\nu}}{g^{n \sigma}(x) \, p_{\sigma}} 
\, ,
\end{equation}
\begin{equation}\label{eq:tHamp}
\frac{d p_{\mu}}{du} \, = \, - \, 
\frac{\tilde{H}(x,p) \, 
(  \partial _{\mu} g^{n \rho}) (x) \, p_{\rho}}{
g^{n \sigma}(x) \, p_{\sigma}}
\, + \, 
\frac{(\partial _{\mu} g^{\rho \tau})(x) \, p_{\rho} \, p_{\tau}}{
2 \, g^{n \sigma}(x) \, p_{\sigma}} 
\, ,
\end{equation}
and the side-condition (\ref{eq:cone}) is equivalent to
\begin{equation}\label{eq:tcone}
\tilde{H}(x,p) \, = \, 0 \, .
\end{equation}  
With (\ref{eq:tcone}), the $\mu = n$ component of (\ref{eq:tHamx})
reduces to
\begin{equation}\label{eq:tHamxn}
\frac{d x^n}{du} \, = \, 1 \, ,
\end{equation}
so the new curve parameter $u$ coincides, indeed, with the coordinate
$x^n$ (up to an additive constant that can be chosen at will). The other 
components of (\ref{eq:tHamx}) take the form
\begin{equation}\label{eq:tHamxi}
\frac{d x^i}{du} \, = \, 
\frac{g^{i \nu} (x) \, p_{\nu}}{g^{n \sigma}(x) \, p_{\sigma}} 
\, ,
\end{equation}
where, as before, latin indices range from 1 to $(n-1)$, and 
(\ref{eq:tHamp}) reduces to
\begin{equation}\label{eq:tHampmu}
\frac{d p_{\mu}}{du} \, = \, 
\frac{(\partial _{\mu} g^{\rho \tau})(x) \, p_{\rho} \, p_{\tau}}{
2 \, g^{n \sigma}(x) \, p_{\sigma}} 
\, .
\end{equation}
Applying the derivative operator $d/du$ to (\ref{eq:tHamxi}), and
using (\ref{eq:tHampmu}), reproduces the equation (\ref{eq:geored})
for $u$-parametrized geodesics. We have thus proven that the 
lightlike geodesics of the metric $g = g_{\mu \nu 
} (x) d x^{\mu} dx^{\nu}$, if reparametrized by $u = x^n$,
can be derived from the Hamiltonian (\ref{eq:Hamred}) together
with the side-condition (\ref{eq:tcone}). 

This gives us a convenient method of how to derive the generalized
Jacobi equation for lightlike geodesics parametrized by $u = x^n$:
Start out from the Hamiltonian (\ref{eq:Hamred}). Write Hamilton's
equations with the side-condition (\ref{eq:tcone}). Rewrite this
as second-order equations for the $x^{\mu}$. Fix a solution
$X^{\mu} (u)$ to these equations, substitute $x^{\mu} (u) = 
X^{\mu} (u) + \xi ^{\mu} (u)$, and linearize with respect to the 
$\xi ^{\mu}$ but not with respect to the $d \xi ^{\mu} /du$.
 
The resulting set of equations describes lightlike geodesics $X(u)+ \xi (u)$
which are close to the lightlike geodesic $X(u)$ but whose tangent 
vectors need not be close to the tangent vector of $X(u)$. For 
$n=4$ these equations can be used, by choosing initial conditions 
appropriately, for describing a homocentric light bundle around a 
central light ray $X(u)$ in a general-relativistic spacetime. If the
vertex of the bundle is at $u=u_0$, say, we have to choose the
initial conditions as $\xi ^i (u_0) = 0$. There are only $(n-2)=2$
independent solutions because the side-condition
(\ref{eq:lightlike}) fixes one of the $d \xi ^i /du$ in terms of the
others. This is in agreement with the intuitively obvious fact 
that a homocentric light bundle has $(n-2)=2$ dimensions transverse 
to the propagation direction. In contrast to the ordinary 
Jacobi equation, the generalized Jacobi equation can be used
to describe homocentric bundles whose opening angle is arbitrarily 
large but the approximation is valid, in general, only for small 
values of $u-u_0$, i.e., close to the vertex of the bundle.
For larger values of $u-u_0$, the $\xi ^i (u)$ will, in general,
not be small, so the fact that we linearized with respect to these
quantities may produce large errors.

To be sure, there are special examples were the generalized Jacobi 
equation holds for a large parameter interval. An example of this 
kind will be given in Section \ref{sec:Fermi}; in this example 
even the Jacobi equation, the generalized 
Jacobi equation and the exact geodesic equation coincide. In general, 
however, the generalized Jacobi equation for lightlike geodesics is 
a short-time equation, describing the temporal evolution of light 
bundles with arbitrarily large opening angles near their vertex. 

We illustrate the general results of this section with an example.

\noindent
{\bf Example 1}: We want to calculate, with the help
of the generalized Jacobi equation, the evolution of
a light bundle around a circular geodesic at $r=3m$
in Schwarzschild spacetime. The Schwarzschild metric
is
\begin{equation}\label{eq:gschw}
g _{\mu \nu} \, dx^{\mu} \, dx ^{\nu} \, = \, - \, 
\big( 1- \frac{2m}{r} \big) \, dt^2 \, + \frac{dr^2}{1- \frac{2m}{r}}
\, + \, r^2 \, d \vartheta ^2 \, + \, r^2 \, \mathrm{sin} ^2 \vartheta \,
d \varphi ^2 \, .
\end{equation}
It is well-known that a lightlike geodesic that starts tangentially
to the circle $r=3m$, $\vartheta = \pi /2$, will stay on this circle.
We want to write the generalized Jacobi equation for lightlike geodesics
near this circular geodesic. We will use the azimuthal coordinate
$\varphi$ for the parameter, $u = x^n = \varphi$. This excludes geodesics
tangent to a half-space $\varphi = \mathrm{constant}$. Therefore, 
our parametrization allows us to treat bundles with any opening 
angle smaller than $\pi /2$ around the circular geodesic, but not 
the limiting case that the opening angle is equal to $\pi /2$. 
Using $\varphi$ for the parameter gives us directly the intersection 
of the bundle with any half-plane $\varphi = \mathrm{constant}$.

After calculating the contravariant components $g^{\mu \nu}$ of
the Schwarzschild metric, we write the Hamiltonian (\ref{eq:Hamred}) 
with $x^n = \varphi \,$,
\begin{equation}\label{eq:tschw}
\tilde{H} (x,p) \, = \, \frac{1}{2} \, 
\left(
\, p_{\varphi} \, + \, 
\frac{\mathrm{sin}^2 \vartheta}{p_{\varphi}} \, 
\Big(
\, p_{\vartheta}^2
\, + \, 
r \, ( r-2m) \, p_r^2 
\, - \,
\frac{r^3\, p_t^2}{r-2m}
\,
\Big)
\, \right) \, .
\end{equation}
Using the side-condition $\tilde{H}(x,p) = 0$, Hamilton's equations
for the Hamiltonian $\tilde{H}$ take the form 
\begin{gather}\label{eq:schwx}
\frac{d \varphi}{d u} \, = \, 1 \, , \quad 
\frac{d t}{d u} \, = \, - \, 
\frac{r^3 \mathrm{sin}^2 \vartheta \, p_t
}{
(r-2m) \, p_{\varphi}} \, , \quad 
\frac{d \vartheta}{d u} \, = \, 
 \frac{\mathrm{sin}^2 \vartheta \, p_{\vartheta}
}{ p_{\varphi}} \, , 
\\[0.3cm]
\nonumber
\frac{d r}{d u} \, = \, 
\frac{r \, (r-2m) \, \mathrm{sin}^2 \vartheta \, p_r
}{p_{\varphi}} \, , 
\end{gather}
\begin{gather}\label{eq:schwp}
\frac{d p_{\varphi}}{d u} \, = \, 0 \, , \quad 
\frac{d p_t}{d u} \, = \, 0  \, , \quad 
\frac{d p_{\vartheta}}{d u} \, = \, 
\frac{\mathrm{cos} \, \vartheta \, p_{\varphi}
}{ \mathrm{sin} \vartheta} \, ,
\\[0.3cm]
\nonumber 
\frac{d p_r}{d u} \, = \, 
- \, \frac{(r-m) \, \mathrm{sin}^2 \vartheta \, p_r^2
}{p_{\varphi}} \, + \,
\frac{r^2 (r-3m) \, \mathrm{sin}^2 \vartheta \, p_t^2
}{(r-2m)^2 \, p_{\varphi}} \, .  
\end{gather}
If we apply the derivative $d/d u$ to the expressions
for $d \vartheta /d u$ and $dr / d u$ from
(\ref{eq:schwx}), use (\ref{eq:schwp}) and the side-condition
$\tilde{H} (x,p) = 0$, we arrive at the following second order 
system for $\vartheta ( u )$ and $r(u)$.
\begin{equation}\label{eq:theta}
\frac{d^2 \vartheta}{d u ^2} \, = \, 
\frac{2 \, \mathrm{cos} \, \vartheta}{\mathrm{sin} \, \vartheta}
\, \Big( \, \frac{d \vartheta}{d u} \, \Big) ^2 
\, + \, 
\mathrm{sin}\, \vartheta \, \mathrm{cos} \, \vartheta
\, .
\end{equation}
\begin{equation}\label{eq:r}
\frac{d^2 r}{d u ^2} \, = \, 
\frac{2 \, \mathrm{cos} \, \vartheta}{\mathrm{sin} \, \vartheta}
\, \frac{d \vartheta}{d u} \,  \frac{d r}{d u}  
\, + \, 
\frac{\, 2 \,}{r} \, 
\Big( \, \frac{d r}{d u} \, \Big) ^2 
\, + \, 
( r-3m) \, 
\left( 
\, \Big( \, \frac{d \vartheta}{d u} \, \Big) ^2 
\, + \, \mathrm{sin}^2 \vartheta \, 
\right)
\, .
\end{equation}
Obviously, this system of equations admits the solution
\begin{equation}\label{eq:circ}
r( u ) \, = \, 3 \, m \, , \qquad 
\vartheta ( u ) \, = \, \frac{\pi}{2} \, .
\end{equation}
In order to linearize around this circular lightlike 
geodesic, we write
\begin{equation}\label{eq:xi}
r ( u ) \, = \, 3 \, m \, + \, 
\frac{1}{\sqrt{3}} \, \xi ^r (u) \, , \qquad 
\vartheta ( u ) \, = \, \frac{\pi}{2} \, + \, 
\frac{1}{3 \, m} \, \xi ^{\vartheta} (u) \, . 
\end{equation}
The numerical factors are chosen such that
\begin{equation}\label{eq:norm}
g( \partial _{\xi ^r} ,  \partial _{\xi ^r} )
\, = \,  
g( \partial _{\xi ^{\vartheta}} ,  \partial _{\xi ^{\vartheta}} ) 
\, = \, 1 
\end{equation}
at $r=3m$, $\vartheta = \pi /2$.  
Inserting (\ref{eq:xi}) into (\ref{eq:theta}) and
(\ref{eq:r}), and linearizing with respect to 
$\xi ^{\vartheta}$ and $\xi ^r$, but not with
respect to $d \xi ^{\vartheta} / d u$ 
and $d \xi ^r / d u$, gives us the 
generalized Jacobi equation:
\begin{equation}\label{eq:xitheta}
\frac{d^2 \xi ^{\vartheta}}{d u ^2}
\, = \, 
- \, \xi ^{\vartheta} \, - \, 
\frac{2 \, \xi ^{\vartheta}}{9 \, m^2} \,
\, \Big( \, \frac{d \xi ^{\vartheta}}{d u} \, \Big) ^2 
\,  ,
\end{equation}
\begin{gather}\label{eq:xir}
\frac{d^2 \xi ^r}{d u ^2}
\, = \, \xi ^r \, + \, 
\frac{\xi ^r}{9 \, m^2} \, 
\left( \, 
\Big( \, \frac{d \xi ^{\vartheta}}{d u} \, \Big) ^2 
\,  - \, 
\frac{2}{3} \,
\, \Big( \, \frac{d \xi ^r}{d u} \, \Big) ^2
\, \right)
\\[0.3cm]
\nonumber
 - \, 
\frac{2 \, \xi ^{\vartheta}}{9 \, m^2} \, 
\frac{d \xi ^{\vartheta}}{d u} \, \frac{d \xi ^r}{du}
\, + \, 
\frac{2}{3 \, \sqrt{3} \, m} \, 
\Big( \, \frac{d \xi ^r}{d u} \, \Big) ^2 
\, .
\end{gather}
If we linearize also with respect to $d \xi ^{\vartheta} / d u$ 
and $d \xi ^r / d u$, only the first term on the right-hand side
of (\ref{eq:xitheta}) and of (\ref{eq:xir}) survives,
\begin{equation}\label{eq:xithetalin}
\frac{d^2 \xi ^{\vartheta}}{d u ^2}
\, = \, 
- \, \xi ^{\vartheta} \,  ,
\end{equation}
\begin{equation}\label{eq:xirlin}
\frac{d^2 \xi ^r}{d u ^2}
\, = \, \xi ^r \, .
\end{equation}
This is the ordinary Jacobi equation. 

\begin{figure}
\begin{center}
   \psfrag{a}{\large $\xi ^r$}  
   \psfrag{b}{\large $\xi ^{\vartheta}$}  
  \includegraphics[width=10cm]{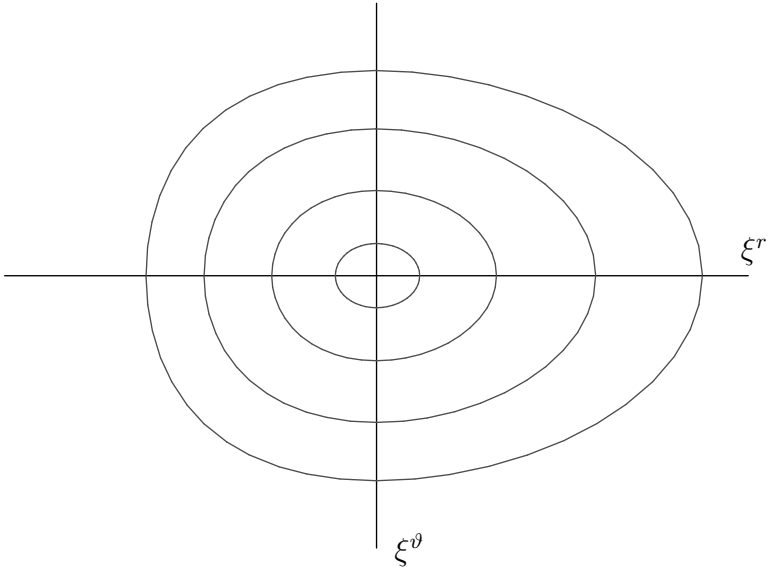}
\end{center}
\caption{Cross section of initially circular light bundle around
a circular lightlike geodesic in Schwarzschild spacetime,
calculated with the generalized Jacobi equation. The point 
$(\xi ^r , \xi ^{\vartheta} ) = (0, 0)$ represents the circular
geodesic at $r=3m$ and $\vartheta = \pi /2$. The $\xi ^r$ axis
is in the equatorial plane, pointing outwards. The $\xi ^{\vartheta}$
axis is perpendicular to the equatorial plane, pointing downwards.
The picture shows the intersection with a half-plane $\varphi =
\mathrm{constant}$ close to the vertex of the bundle, for four 
different opening angles.
}
\label{fig:schw}       
\end{figure}

If we solve  (\ref{eq:xithetalin}) and (\ref{eq:xirlin}) 
with initial conditions
\begin{equation}\label{eq:initial1}
\xi ^{\vartheta} (0) \, = \, 0 \, , \quad \xi ^r (0) \, = \, 0 \, , 
\end{equation}
\begin{equation}\label{eq:initial2}
\frac{d \xi ^{\vartheta}}{d \varphi} (0) \, = \, 
- \, \varepsilon \, \mathrm{sin}\, \chi \, , \quad 
\frac{d \xi ^r}{d \varphi} (0) \, = \, 
 \varepsilon \, \mathrm{cos}\, \chi \, , \quad
\end{equation}
with $\chi$ running from 0 to $2 \pi$ and $\varepsilon$ fixed, it 
gives us the cross-section of an initially circular bundle with 
opening angle proportional to $\varepsilon$. Owing to the linearity
of the ordinary Jacobi equation, the cross-section of such a bundle 
will be elliptic for all values of $u = \varphi$. As a consequence 
of the plus sign on the right-hand side of (\ref{eq:xirlin}), in
contrast to the minus sign on the right-hand side of (\ref{eq:xithetalin}),
the expansion of the bundle will increase in the $\xi ^r$ direction
and decrease in the $\xi ^{\vartheta}$ direction; hence the major 
axis of the ellipse is in the $\xi ^r$ direction. This reflects 
the fact that the circular geodesic at $r=3m$ is unstable with 
respect to perturbations in the $\xi ^r$ direction but stable 
with respect to perturbations in the $\xi ^{\vartheta}$ direction. 

By contrast, if we solve the non-linear equations (\ref{eq:xitheta})
and (\ref{eq:xir}) with initial conditions (\ref{eq:initial1})
and (\ref{eq:initial2}), the cross-section of the resulting
bundle will not be elliptic. The larger the opening angle $\varepsilon$,
the stronger the deviation from the elliptic shape due to 
the non-linearities, see Figure \ref{fig:schw}. 

This example demonstrates how the generalized Jacobi equation 
can be used for calculating the shapes of light bundles, beyond
the small-angle approximation that is inherent in the standard
treatment based on the ordinary Jacobi equation. Of course,
one has to keep in mind that the generalized Jacobi equation
is, in genetral, a valid approximation only close to the vertex 
of the bundle. 

\section{The generalized Jacobi equation for lightlike geodesics in
Fermi coordinates}
\label{sec:Fermi}
We have already discussed how, on an affine manifold, Fermi coordinates
can be introduced near a reference geodesic $X(s)$. The construction involved
the choice of a parallely transported $n$-bein $E_1 (s), \dots , E_n (s)$
along the chosen geodesic which was arbitrary apart from the fact that 
$E_n (s)$ should coincide with the tangent vector $dX(s)/ds$ of the 
reference geodesic. If our connection is the Levi-Civita connection of a 
Lorentzian metric, we can further specify this $n$-bein. For a timelike
reference geodesic, it is usual to require that the $n$-bein be orthonormal. 
Then $E_1 (s), \dots , E_{n-1} (s)$ span the spacelike orthocomplement of 
$dX(s)/ds$ at each value of $s$. Thus, in the Fermi coordinates 
$(x^1, \dots , x^{n-1},u)$ the hypersurfaces $u = \mathrm{constant}$ 
intersect the timelike reference geodesic $X(s)$ orthogonally and are, 
therefore, spacelike near X(s). (Farther away from $X(s)$ they need not 
be spacelike.) These are the Fermi normal coordinates treated in 
standard text-books on general relativity. In these coordinates the
metric attains a standard form if written up to second order in the
transverse coordinates. This standard form was derived by Manasse
and Misner \cite{ManasseMisner1963} and can be found, e.g., in 
Misner, Thorne and Wheeler \cite{MisnerThorneWheeler1973}, p.332.

Whereas this standard text-book treatment of Fermi normal coordinates
assumes a timelike geodesic, here we are interrested in Fermi coordinates 
near a lightlike reference geodesic $X(s)$. Then the $n$-bein cannot be 
chosen orthonormal because of the requirement that $E_n (s) = dX(s)/ds$. 
The best choice is to have $E_{n-1} (s)$ lightlike with 
\begin{equation}\label{eq:Enorm}
g_{\mu \nu} \big( X(s) \big) \, 
E^{\mu} _{n-1} (s) \, E^{\nu} _n (s) \, = \, - \, 1 \, ,
\end{equation}
and the remaining vectors $E_1 (s) , \dots , E_{n-2} (s)$ 
orthonormal and perpendicular to both $E_{n-1}$ and $E_n$. (Condition
(\ref{eq:Enorm}) assures that $E_{n-1}$ is future-pointing if $E_n$
is future-pointing.) With this choice, the resulting Fermi coordinates 
$(x^1, \dots , x^{n-2}, x^{n-1}=v , u)$ yield hypersurfaces 
$u= \mathrm{constant}$ that are lightlike where they meet the 
reference geodesic $X(s)$. The hypersurface $v=0$ is lightlike along the 
reference geodesic $X(s)$ which is completely contained in this hypersurface. 
These lightlike Fermi normal coordinates were discussed in some detail in a 
recent article by Blau, Frank and Weiss \cite{BlauFrankWeiss2006}. In that 
article, the authors derive the general expression for the metric in lightlike 
Fermi normal coordinates, up to second order in the coordinates away from
the reference geodesic. This result is the lightlike analogue of the above-mentioned
Manasse-Misner representation of the metric near a timelike geodesic. 
If adapted to our sign and index conventions, it reads
\begin{gather}\label{eq:Fermig}
g_{\mu \nu} \, dx^{\mu} \, dx^{\nu} \, = \, 
- \, 2\, du \, dv \, + \, \delta _{AB} \, dx^A \,dx^B \, - \,
R_{injn}(u) \, x^i \, x^j \, du^2 
\\
\nonumber
 - \, 
\frac{4}{3} \, R_{ikjn} (u) \, x^i\, x^j \, dx^k \, du \, - \,  
\frac{1}{3} \, R_{ikjl} (u) \, x^i\, x^j  \, dx^k \, dx^l \, \dots
\end{gather}
Here the Fermi coordinates are denoted $(x^1 , \dots , x^{n-2}, x^{n-1}=v , u)$,
as outlined above. As before, greek indices run from 1 to $n$ and lower
case latin indices $i,j,k,l, \dots$ run from 1 to $(n-1)$. In addition, upper 
case latin indices $A,B, \dots$ run from 1 to $(n-2)$. As usual, $\delta _{AB}$
denotes the Kronecker delta. $R_{\rho \nu \sigma \tau} (u) \, = \, 
g_{\rho \mu} (u) \, R^{\mu} _{\nu \sigma \tau} (u)$ is the purely covariant
version of the curvature tensor (\ref{eq:curv}), evaluated at 
$(x^1=0, \dots , x^{n-2}=0, x^{n-1} = v =0 , u)$, i.e., along the reference 
geodesic. The ellipses in (\ref{eq:Fermig}) indicate terms of third and
higher order with respect to the $x^i$.

With the contravariant metric components $g^{\mu \nu}$ calculated, up to
second order, from (\ref{eq:Fermig}), we can write the Hamiltonian
(\ref{eq:Hamred}) up to second order. This is enough to write the 
pertaining Hamilton equations up to first order, which together with the
constraint (\ref{eq:tcone}) will give us the generalized Jacobi equation 
in lightlike Fermi normal coordinates near an arbitrary lightlike 
reference geodesic. 

We will illustrate this with an example.

\vspace{0.3cm}
\noindent
{\bf Example 2}:
Consider a plane-wave spacetime in Brinkmann coordinates 
$(x^1 , \dots , x^{n-2}, v , u)$,
\begin{equation}\label{eq:plane}
g_{\mu \nu} \, dx^{\mu} \, dx^{\nu} \, = \, 
- \, 2 \, du \, dv \, - \,
h_{AB}(u) \, x^A \, x^B \, du^2  \, + \, 
\delta _{AB} \, dx^A \,dx^B \, ,
\end{equation}
where $h_{AB}(u)= h_{BA}(u)$ has a non-negative trace,
\begin{equation}\label{eq:trace}
\delta ^{AB} \, h_{AB} (u) \, \ge \, 0\, ,
\end{equation}
but is arbitrary otherwise. For $n=4$, any such metric can 
be interpreted as a combined gravitational and electromagnetic 
plane wave. If equality holds in (\ref{eq:trace}), the 
spacetime is Ricci-flat and, thus, a pure gravitational 
wave. For any choice of $h_{AB}(u)$, the vector field
$\partial _v$ is lightlike and absolutely parallel. For a 
discussion of the geometry of plane-wave spacetimes the reader 
is refered to Penrose \cite{Penrose1965}.

Plane-wave spacetimes have the following interesting property,
first discovered by Penrose \cite{Penrose1976}. Near \emph{any}
lightlike geodesic in \emph{any} spacetime, the metric takes
the form of a plane wave in a well-defined limit, called the
Penrose limit. Thereby the original geodesic is represented as the 
curve $(x^1=0, \dots, x^{n-2}=0,v=0,u)$ in the limiting plane-wave
spacetime (\ref{eq:plane}). The Penrose limit can be conveniently 
written in terms of lightlike Fermi normal coordinate, as recently
demonstrated by Blau, Frank and Weiss \cite{BlauFrankWeiss2006}.
In particular, their analysis showed that Brinkmann coordinates
for plane waves are lightlike Fermi normal coordinates; in this case
all the higher-order terms, which are indicated in (\ref{eq:Fermig}) 
by ellipses, vanish exactly. The curvature tensor is given along
the geodesic $(x^1=0, \dots, x^{n-2}=0,v=0,u)$ by
\begin{equation}\label{eq:RBrink}
R_{AnBn}(u) \,  = \, h_{AB} (u) \, 
\end{equation}
and $R_{\mu \nu \sigma \tau }(u) \, = \, 0$ for all other index
combinations, compare (\ref{eq:Fermig}) with (\ref{eq:plane}).

We want to write the generalized Jacobi equation near the 
lightlike geodesic $(x^1=0, \dots , x^{n-2}=0, v=0, u)$.
We will use $u=x^n$ for the curve parameter. After
calculating  the contravariant metric components from (\ref{eq:plane}),
we can write the Hamiltonian (\ref{eq:Hamred}):
\begin{equation}\label{eq:Hamplane}
\tilde{H}(x,p) \, = \, 
 p_u \, - \, \frac{1}{2} \, h_{AB}(u) \, x^A \, x^B \, p_v
\, - \, \frac{\, \delta ^{AB} \, p_A \, p_B \,}{2 \, p_v} \, .
\end{equation}
Note that this Hamiltonian is of second order with respect
to the coordinates $x^A$ and independent of the coordinate $v$.
Hamilton's equations with the Hamiltonian (\ref{eq:Hamplane})
take the form
\begin{equation}\label{eq:xplane}
\frac{du}{du} \, = \, 1 \, , \quad
\frac{dv}{du} \, = \, - \,  
\frac{1}{2} \, h_{AB}  \, x^A \, x^B \, + \, 
 \frac{\delta ^{AB}\, p_A \, p_B}{2 \, p_v^2} \, , \quad
\frac{dx^A}{du} \, = \, - \, \frac{\delta^{AB} \, p_B}{p_v} \, , 
\end{equation}
\begin{equation}\label{eq:pplane}
\frac{dp_u}{du} \, = \,  
\frac{1}{2} \, \frac{dh_{AB}}{du} \,  x^A \, x^B \, p_v \, , \quad
\frac{dp_v}{du} \, = \, 0 \, , \quad
\frac{dp_A}{du} \, = \, h_{AB} \, x^B \, p_v \, . 
\end{equation}

\begin{figure}
\begin{center}
   \psfrag{u}{$v-u$}  
   \psfrag{v}{$v+u$}  
   \psfrag{x}{$x^1$}  
  \includegraphics[width=10.5cm]{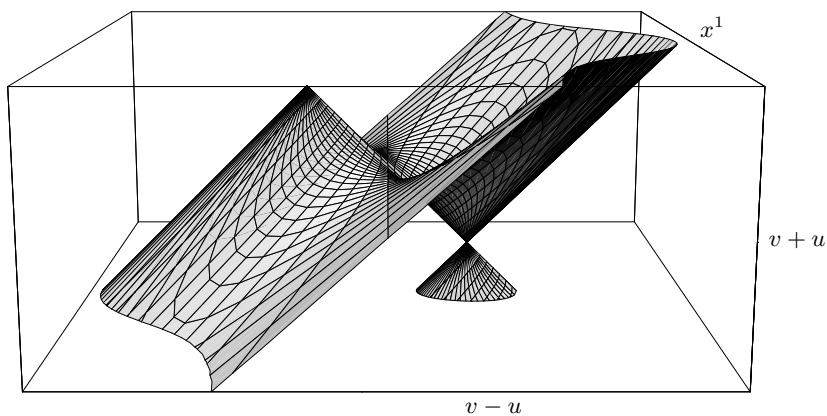}
\end{center}
\caption{Past light cone of an event in a plane-wave spacetime
(\protect\ref{eq:plane}). Three dimensions $(x^1,v,u)$ are
shown, with $u$ future-pointing lightlike (from lower right
to upper left), $v$ future-pointing lightlike (from lower
left to upper right) , and $x^1$ orthogonal to both.
The picture is valid for the case that $x^1$ is an 
eigendirection of $h_{AB}(u)$ with positive eigenvalue. In
this case in the three-dimensional picture all the light rays 
issuing from an event into the past are refocused into another 
event, with the exception of one single light ray that stays 
on an integral curve of the absolutely parallel lightlike 
vector field $\partial _v$. A colour version of this picture 
can be found online in \protect\cite{Perlick2004}. For a similar 
picture, hand-drawn by Roger Penrose, see \protect\cite{Penrose1965}.
}
\label{fig:wave}       
\end{figure}

\noindent
Applying $d/du$ to (\ref{eq:xplane}), and using (\ref{eq:pplane}),
gives us the lightlike geodesic equation in second-order form:
\begin{equation}\label{eq:xA}
\frac{d^2 x^A}{du^2} \, = \, - \, \delta ^{AB} \, h_{BC} \, x^C \, ,
\end{equation}
\begin{equation}\label{eq:v}
\frac{d^2 v}{du^2} \, = \, - \, 
\frac{1}{2} \, \frac{dh_{AB}}{du} \, x^A \, x^B \, - \, 
2 \, h_{AB} \, x^A \, \frac{dx^B}{du} \, .
\end{equation}
If we write 
\begin{equation}\label{eq:xiplane}
x ^A (u) \, = \, 0 \, + \, \xi ^A (u) \, , \quad
v (u) \, = \, 0 \, + \, \xi ^v (u) \, ,
\end{equation}
and linearize with respect to $\xi ^A$ and $\xi ^v$, we
get the generalized Jacobi equation 
\begin{equation}\label{eq:genjacplane}
\frac{d^2 \xi ^A}{du^2} \, = \, - \, 
\delta ^{AB} \, h_{BC} \, \xi ^C \, ,
\quad
\frac{d^2 \xi ^v}{du^2} \, = \, 
 - \, 2 \, h_{AB} \, \xi ^A \, \frac{d \xi ^B}{du}   .
\end{equation}
For the transverse coordinates $x^A = \xi ^A$, these equations are 
independent of the velocities; hence, the generalized Jacobi 
equation coincides with the ordinary geodesic equation. Moreover, 
the (generalized) Jacobi equation even coincides with the exact 
geodesic equation (\ref{eq:xA}). If we solve these equations with 
initial conditions 
\begin{equation}\label{eq:init}
x^A (u_0) = 0 \, , \quad
\delta _{AB} \, \frac{dx^A}{du}(u_0)
\, \frac{dx^B}{du}(u_0) \, = \, \varepsilon ^2 \, ,
\end{equation}
it gives us the cross-section
of an initially circular light bundle around the geodesic
$(x^1 = 0 , \dots , x^{n-2} = 0 , v=0 , u)$ with vertex
at $u=u_0$. Such a bundle will have an elliptic cross-section, 
for arbitrarily large opening angle. The rate of expansion 
is positive in eigen-directions of $h_{AB} (u)$ with negative 
eigenvalues and negative in eigen-directions of $h_{AB} (u)$ 
with positive eigenvalues. Condition (\ref{eq:trace}) 
makes sure that at least one eigenvalue is positive (unless
all are zero in which case (\ref{eq:plane}) is just the
Minkowski metric). In directions with positive eigenvalue the
light bundle will be refocussed until a conjugate point is reached. 
This focusing property is illustrated by Figure \ref{fig:wave} 
which displays the past-light cone of an event in a plane-wave 
spacetime, with three dimensions $(x^1,v,u)$ shown. 

The fact that, for the lightlike geodesic $(x^1=0, \dots, x^{n-2}=0,v=0,u)$
in a plane-wave spacetime, the generalized Jacobi equation coincides with 
the ordinary Jacobi equation has the following interesting consequence:
In the Penrose limit the difference between the generalized Jacobi equation 
and the Jacobi equation vanishes.

\section{Concluding remarks}
\label{sec:conclusion}
In this article we have discussed a generalized Jacobi equation for 
lightlike geodesics, following as closely as possible the approach
that was brought forward by Hodgkinson, Mashhoon and others for timelike 
geodesics. As an alternative, one could also apply the approach
of Ba{\.z}a{\'n}ski \cite{Bazanski1977}, which was mentioned in
the introduction, to lightlike geodesics.
In that case one would consider lightlike geodesics of the form 
$x(s) \, = \, X(s)  \, + \, \epsilon \, \xi (s)$, where $X(s)$ is a lightlike 
reference geodesic, and solve the geodesic equation for $x(s)$ 
iteratively up to some order $N$ with respect to $\epsilon$. 
For any finite order $N$, the resulting equation would not be 
valid for light bundles of arbitrarily large opening angle,
in contrast to the generalized Jacobi equation treated here.
On the other hand, for $\epsilon$ sufficiently small it would be 
valid for arbitrarily large parameter intervals. An interesting
application of this Ba{\.z}a{\'n}ski-type approach to lightlike
geodesics, which apparently has not been considered in the 
literature so far, could be to study the caustics of light
bundles up to some order $N$ with respect to $\epsilon$.


\bibliographystyle{spmpsci}      



\begin{thebibliography}{10}
\providecommand{\url}[1]{{#1}}
\providecommand{\urlprefix}{URL }
\expandafter\ifx\csname urlstyle\endcsname\relax
  \providecommand{\doi}[1]{DOI~\discretionary{}{}{}#1}\else
  \providecommand{\doi}{DOI~\discretionary{}{}{}\begingroup
  \urlstyle{rm}\Url}\fi

\bibitem{Bazanski1977}
Ba{\.z}a{\'n}ski, S.: Kinematics of relative motion of test particles in
  general relativity.
\newblock Ann. Inst. H. Poincar{\'e} (A) \textbf{27}, 115--144 (1977)

\bibitem{BlauFrankWeiss2006}
Blau, M., Frank, D., Weiss, S.: {F}ermi coordinates and {P}enrose limit.
\newblock Class. Quant. Grav. \textbf{23}, 3993--4010 (2006)

\bibitem{ChiconeMashhoon2002}
Chicone, C., Mashhoon, B.: The generalized {J}acobi equation.
\newblock Class. Quant. Grav. \textbf{19}, 4231--4248 (2002)

\bibitem{ChiconeMashhoon2005}
Chicone, C., Mashhoon, B.: Ultrarelativistic motion: inertial and tidal effects
  in {F}ermi coordinates.
\newblock Class. Quant. Grav. \textbf{22}, 195--205 (2005)

\bibitem{ChiconeMashhoon2006}
Chicone, C., Mashhoon, B.: Explicit {F}ermi coordinates and tidal dynamics in
  de {S}itter and {G}{\"o}del spacetimes.
\newblock Phys. Rev. \textbf{D 74}, 064,019 (2006)

\bibitem{Ciufolini1986}
Ciufolini, I.: Generalized geodesic deviation equation.
\newblock Phys. Rev. \textbf{D 34}, 1014--1017 (1986)

\bibitem{ColisteteLeygnacKerner2002}
Colistete, R., Leygnac, C., Kerner, R.: Higher-order geodesic deviations
  applied to the {K}err metric.
\newblock Class. Quant. Grav. \textbf{19}, 4573--4589 (2002)

\bibitem{HawkingEllis1973}
Hawking, S.W., Ellis, G.F.R.: The large scale structure of space-time.
\newblock Cambridge University Press, Cambridge (1973)

\bibitem{Hodgkinson1972}
Hodgkinson, D.E.: A modified equation of geodesic deviation.
\newblock Gen. Relativ. Gravit. \textbf{3}, 351--375 (1972)

\bibitem{KernerHoltenColistete2001}
Kerner, R., van Holten, J.W., Colistete R., J.: Relativistic epicycles: another
  approach to geodesic deviations.
\newblock Class. Quant. Grav. \textbf{18}, 4725--4742 (2001)

\bibitem{ManasseMisner1963}
Manasse, F.K., Misner, C.W.: Fermi normal coordinates and some basic concepts
  in differential geometry.
\newblock J. Math. Phys. \textbf{4}, 735--745 (1963)

\bibitem{Mashhoon1975}
Mashhoon, B.: On tidal phenomena in a strong gravitational field.
\newblock Astrophys. J. \textbf{197}, 705--716 (1975)

\bibitem{Mashhoon1977}
Mashhoon, B.: Tidal radiation.
\newblock Astrophys. J. \textbf{216}, 591--609 (1977)

\bibitem{MisnerThorneWheeler1973}
Misner, C., Thorne, K., Wheeler, J.A.: Gravitation.
\newblock Freeman, San Francisco (1973)

\bibitem{Penrose1965}
Penrose, R.: A remarkable property of plane waves in general relativity.
\newblock Rev. Modern Phys. \textbf{37}, 215--220 (1965)

\bibitem{Penrose1976}
Penrose, R.: Any space-time has a plane wave as limit.
\newblock In: M.~Cahen, M.~Flato (eds.) Differential geometry and relativity,
  pp. 271--275. Reidel, Dordrecht (1976)

\bibitem{Perlick2004}
Perlick, V.: Gravitational lensing from a spacetime perspective.
\newblock Living Rev. Relativity \textbf{7(9)} (2004).
\newblock {h}ttp://www.livingreviews.org/lrr-2004-9

\bibitem{SwaminarayanSafko1983}
Swaminarayan, N.S., Safko, J.L.: A coordinate-free derivation of a generalized
  geodesic deviation equation.
\newblock J. Math. Phys. \textbf{24}, 883--885 (1983)

\bibitem{Synge1960}
Synge, J.L.: {Relativity. The general theory}.
\newblock North-Holland, Amsterdam (1960)

\end{thebibliography}

%
%

\end{document}